\documentclass[thmsa,twocolumn]{article}
\usepackage{sw20lart}

\input{tcilatex}
\begin{document}

\author{\textbf{C. Figueira de Morisson Faria}$^{\dagger }$\textbf{, A. Fring}$%
^{\ddagger }$\textbf{\ and R. Schrader}$^{\ddagger }$ \\
$\dagger $Max-Born-Institut, Rudower Chaussee 6, D-12474 Berlin, Germany\\
$\ddagger $Institut f\"{u}r Theoretische Physik, Freie Universit\"{a}t
Berlin, \\
Arnimallee 14, D-14195 Berlin, Germany}
\title{Analytical Treatment of Stabilization}
\date{August 1998}
\maketitle

\begin{abstract}
We present a summarizing account of a series of investigations whose central
topic is to address the question whether atomic stabilization exists in an
analytical way. We provide new aspects on several issues of the matter in
the theoretical context when the dynamics is described by the Stark
Hamiltonian. The main outcome of these studies is that the governing
parameters for this phenomenon are the total classical momentum transfer and
the total classical displacement. Whenever these two quantities vanish,
asymptotically weak stabilization does exist. For all other situations we
did not find any evidence for stabilization. We found no evidence that
strong stabilization might occur. Our results agree qualitatively with the
existing experimental findings.
\end{abstract}

\section{Introduction}

\footnotetext{
e-mail addresses:
\par
\begin{tabular}{l}
Faria@mbi-berlin.de, \\ 
Fring@physik.fu-berlin.de, \\ 
Schrader@physik.fu-berlin.de.
\end{tabular}
}Due to the breakdown of standard perturbation theory, the understanding of
the physics of an atom in a strong (intensities larger than $3.5\times
10^{16}Wcm^{-2}$ for typical frequencies) laser field is still poorly
understood to a very large extent. Hitherto the large majority of the
obtained results is based on numerical treatments. In our investigations we
aim at a rigorous analytical description of phenomena occurring in this
regime. This proceeding will provide an account of a series of publications
[1-4]. In many cases we will simply summarize and state some of the results
and refer the reader for a detailed derivation to the original manuscripts,
but we shall try to put an emphasis on new aspects for which we will supply
an extensive discussion. Several of the presented arguments and results may
not be found in [1-4].

Amongst the phenomena occurring in the high intensity regime in particular
the one of so-called stabilization has recently caused some controversy, not
only concerning its definition, but even its very existence altogether
[5-26]. Roughly speaking stabilization means that atomic bound states become
resistent to ionization in ultra-intense laser fields. A more precise
definition may be found in section 2.2.

\section{Physical Framework}

The object of our investigations is an atom in the presence of a
sufficiently intense\footnote{%
Sufficiently refers to the validity of a classical treatment of the laser
field. A rigerous quantum electrodynamical treatment of ionization phenomena
was recently initiated in \cite{Wu}.} laser field, which may be described in
the non-relativistic regime by the time-dependent Schr\"{o}dinger equation
in the dipole approximation 
\begin{equation}
i\frac{\partial \left| \psi (t)\right\rangle }{\partial t}=H\left( t\right)
\left| \psi (t)\right\rangle .  \label{Schrodi}
\end{equation}
We use atomic units throughout. The time dependent external electric field
will be treated classically and is assumed to be linearly polarized of the
general form $E(t)=E_{0}f(t),$ where $E_{0}$ denotes the field amplitude and 
$f(t)$ is some arbitrary function which equals zero for $t<0$ and $t>\tau ,$
such that $\tau $ defines the pulse length. Depending on the context it is
convenient to express the Hamiltonian in equation (\ref{Schrodi}) in
different gauges.

\subsection{Gauge equivalent Hamiltonians}

Taking $A_{j\leftarrow i}(t)$ to be a one parameter family of unitary
operators, we may construct the gauge equivalent Hamiltonian $H_{i}\left(
t\right) $ from $H_{j}\left( t\right) $ by the usual gauge transformation 
\begin{equation}
H_{i}(t)=i\partial _{t}A_{j\leftarrow i}(t)A_{j\leftarrow
i}^{-1}(t)+A_{j\leftarrow i}(t)H_{j}(t)A_{j\leftarrow i}^{-1}(t).
\label{gauge}
\end{equation}
Choosing the most conventional gauge, the so-called length gauge, the
Hamiltonian to describe the above mentioned physical situation is the Stark
Hamiltonian 
\begin{equation}
H_{l}^{S}(t)=H_{l}^{0}+V\left( \mathbf{\vec{x}}\right) +\mathbf{z}\cdot
E(t)\,.
\end{equation}
$V\left( \mathbf{\vec{x}}\right) $ is the atomic potential and $H_{l}^{0}=%
\mathbf{\vec{p}}^{2}/2$ denotes the Hamilton operator of the free particle.
We introduced here sub-and superscripts in order to keep track of the
particular gauge we are in and to identify a specific Hamiltonian,
respectively. In our conventions $\mathbf{\vec{p}}$ and $\mathbf{\vec{x}}$
denote operators, whilst $\vec{p}$ and $\vec{x}$ are elements in $R^{3}$.
Other commonly used Hamiltonians are the one in the velocity gauge 
\begin{equation}
H_{v}^{S}(t)={\frac{1}{2}}(\mathbf{\vec{p}}-b(t)e_{z})^{2}+V\left( \mathbf{%
\vec{x}}\right)
\end{equation}
and the one in the Kramers-Henneberger (KH) frame \cite{KH} 
\begin{equation}
H_{KH}^{S}(t)=H_{l}^{0}+V(\mathbf{\vec{x}}-c(t)e_{z}).
\end{equation}
Here $e_{z}$ denotes the unit vector in the z-direction. These Hamiltonians
may be obtained from each other by using 
\begin{eqnarray}
A_{v\leftarrow l}(t) &=&e^{ib(t)\mathbf{z}},  \label{g1} \\
A_{v\leftarrow KH}(t) &=&e^{-ia(t)}e^{ic(t)\mathbf{p}_{z}},  \label{g2} \\
A_{l\leftarrow KH}(t) &=&e^{-ia(t)}e^{-ib(t)\mathbf{z}}e^{ic(t)\mathbf{p}%
_{z}},  \label{g3}
\end{eqnarray}
in (\ref{gauge}).\textbf{\ }$\mathbf{p}_{z}$ is the component of the
momentum operator in the z-direction. We have employed the important
quantities 
\begin{eqnarray}
b\left( t\right) &=&\int\limits_{0}^{t}ds\,E\left( s\right) , \\
c\left( t\right) &=&\int\limits_{0}^{t}ds\,b\left( s\right) , \\
a\left( t\right) &=&\frac{1}{2}\int\limits_{0}^{t}ds\,b^{2}\left( s\right) ,
\end{eqnarray}
which are the classical momentum transfer, the classical displacement and
the classical energy transfer, respectively. It will turn out, that in
particular $b\left( \tau \right) $ and $c\left( \tau \right) $ are the
crucial parameters for the description of the phenomenon we are going to
discuss. The classical energy transfer $a(t)$ is not a crucial quantity
since it enters all expressions only as a phase and will therefore cancel in
all relevant physical expressions.

In our considerations we will also need the Hamiltonians 
\begin{equation}
H_{l}^{A}(t)=H_{l}^{0}+V\left( \mathbf{\vec{x}}\right) ,\quad
\,H_{l}^{GV}(t)=H_{l}^{0}+\mathbf{z}\cdot E(t)\,
\end{equation}
which describe an electron in the atomic potential or in the electric field,
respectively. Of course these Hamiltonians may also be transformed into the
other gauges by (\ref{g1})-(\ref{g3}). Notice that $H_{KH}^{GV}(t)=H_{l}^{0}$%
.

\subsection{Definition of Stabilization}

Since stabilization means different things to different authors and a
universally accepted concept does not seem to exist yet, we will precisely
state our definition of it. We will not discuss the behaviour of ionization
rates as some authors do, but we shall consider exclusively ionization
probabilities. Denoting by $\Vert \psi \Vert ^{2}=\langle \psi ,\psi \rangle
=\int |\psi (\vec{x})|^{2}d^{3}x$ the usual Hilbert space norm, the
ionization probability is defined as 
\begin{equation}
\mathcal{P}(\psi )\;=\;1-\Vert P_{+}S\psi \Vert ^{2}\;\;\;.  \label{ion1}
\end{equation}
We used the scattering matrix 
\begin{equation}
S=\lim_{t_{\pm }\rightarrow \pm \infty }\exp (it_{+}H_{+})\cdot
U(t_{+},t_{-})\cdot \exp (-it_{-}H_{-})\;,\;
\end{equation}
where $H_{\pm }=\lim_{t\rightarrow \pm \infty }H(t)$, $\psi $ is a
normalized bound state of $H_{-}$, $P_{\pm }$ is the projector onto the
bound state space of $H_{\pm }$ and $U(t,t^{\prime })$ is the time evolution
operator from time $t^{\prime }$ to $t$ associated\footnote{%
Associated is to be understood in the sense that the time evolution operator
obeys the Schr\"{o}dinger equation $i\partial _{t}U(t,t^{^{\prime
}})=H(t)U(t,t^{^{\prime }})\,.$} to $H(t)$. The time evolution operators may
be transformed from one gauge to another by 
\begin{equation}
U_{i}^{a}(t,t^{\prime })=A_{j\leftarrow i}(t)U_{j}^{a}(t,t^{\prime
})A_{j\leftarrow i}^{-1}(t^{\prime }).  \label{Ugauge}
\end{equation}
The ionization probability $\mathcal{P}(\psi )$ is a gauge invariant
quantity \cite{FKS}. Note that for the gauge equivalent Hamiltonians quoted
above we have in general 
\begin{equation}
\lim_{t\rightarrow \pm \infty }H_{l}^{S}(t)\neq \lim_{t\rightarrow \pm
\infty }H_{v}^{S}(t)\neq \lim_{t\rightarrow \pm \infty }H_{KH}^{S}(t).
\label{limh}
\end{equation}
However, (recall that $b\left( 0\right) =c\left( 0\right) =0$), equality in
the first case holds whenever we have $b\left( \tau \right) =0$ and in both
cases when in addition $c\left( \tau \right) =0$. We will encounter this
condition of a particularly switched on and off pulse below once more as the
necessary condition for the presence of what we refer to as asymptotically
weak stabilization \footnote{%
There are doubts expressed by experimentalists about the possibility to
realise pulses having simultaneously $b\left( \tau \right) =0$ and $c\left(
\tau \right) =0$ \cite{Sauer}.}. We would like to point out that this
condition does not coincide necessarily with the notion of adiabatically
switched on and off pulses, because we may achieve $b\left( \tau \right) =0$
and $c\left( \tau \right) =0$ of course also with a very rapid switch on and
off. Since we are interested in the behaviour of the atomic bound states $%
\left| \psi (t=0)\right\rangle =\left| \psi \right\rangle $ of the
Hamiltonian $H_{l}^{A}$ we should commence the discussion in (\ref{ion1}) in
the length gauge (in this case we have $\lim_{t\rightarrow \pm \infty
}H_{l}^{S}(t)=H_{l}^{A}$) such that in our situation 
\begin{equation}
\mathcal{P}(\psi )=1-\left\| P_{+}U_{l}^{S}\left( \tau ,0\right) \psi
\right\| ^{2}\quad .  \label{ion}
\end{equation}
Regarding the ionization probability as a function of the field amplitude $%
E_{0}$, stabilization means that 
\begin{equation}
\frac{d\mathcal{P}\left( \psi \right) }{dE_{0}}\leq 0\qquad \text{for\quad }%
\mathcal{P}\left( \psi \right) \neq 1  \label{stabilization}
\end{equation}
for $E_{0}\in \left[ 0,\infty \right) $ on a finite interval. Hence the
occurence of a saddle point does not qualify as stabilization. Also we would
like to introduce some terminology in order to distinguish in (\ref
{stabilization}) between the case of equality and strict inequality. If the
former sign holds we call this behaviour \textit{``weak stabilization''} and
in the latter case \textit{``strong stabilization''}. In case weak
stabilization only occurs in the limit $E_{0}\rightarrow \infty ,$ we shall
refer to it as \textit{``asymptotically weak stabilization''.}

\section{Upper and lower bounds for the Ionization Probability}

The outcome of every theoretical investigation will attach some sort of
error to any physical quantity. In the minority of cases this error can be
precisely stated, since it may either be the consequence of various
qualitative assumptions based on some physical reasoning which are difficult
to quantify or it may be of a more technical nature originating in the
method used. For instance for the physical quantity we are interested in,
the ionization probability $P(\psi )$, the most fundamental error is
introduced by the assumptions for the validity of the main physical
framework, that is the Schr\"{o}dinger equation (\ref{Schrodi}) (i.e.
non-relativity, dipole approximation, classical treatment of the external
field, neglect of the magnetic field, etc.). Examples for errors rooted in a
particular method used are: When solving the Schr\"{o}dinger equation
numerically one is forced to discretise $H(t)$, insert the atom into a
finite box and introduce absorbing mask functions at the boundary, etc. Also
one is not able to project on all bound states or all states of the
``discrete continuum''\footnote{%
See the conclusion for a discussion of this point.} and is forced to
introduce a cut-off, whose effect is in our opinion not discussed in the
literature. Some further examples are the errors resulting from the
termination of a Floquet, Fourier expansion or perturbation series\footnote{%
See section 6 and the conclusion for a discussion of this point.}. We would
like the reader to keep these basic facts in mind, i.e. ``exact'' results do
not exist and one is always dealing with some form of bounds, when judging
about the method presented in this section. The essence of the method
consists in treating bounds which restrict a physical quantity rather than
looking at its actual value. One of the main virtues of this approach is
that it may be carried out purely analytically. In different contexts it has
turned out to be extremely fruitful, for instance in the proof of the
stability of matter \cite{Lieb1} and the stability of matter in a magnetic
field \cite{Lieb2}.

We will provide rigorous analytic expressions for the upper and lower bound, 
$P_{u}(\psi )$ and $P_{l}(\psi ),$ respectively, for the ionization
probability in the sense that 
\begin{equation}
P_{l}(\psi )\leq P(\psi )\leq P_{u}(\psi )\,.
\end{equation}
Hence within the basic theoretical framework upper and lower bounds serve as
sharp error bars. Surely one should treat these expressions with care and be
aware of their limitations in the sense that about the actual shape of $%
P(\psi )$ no decisive conclusion can be drawn whenever $P_{l}(\psi )$
differs strongly from $P_{u}(\psi )$. However, it seems a reasonable
assumption that the analytic expression of the bounds reflect qualitatively
the behaviour of the precise ionization probability. Nonetheless, there
exist certain type of questions in the present context which can be answered
decisively with this method. Concerning the question of stabilization we may
consider the bounds as functions of the field amplitude and can conclude
that stabilization exists or does not exist once we find that $P_{u}(\psi )$
for increasing field amplitude tends to zero and $P_{l}(\psi )$ tends to
one, respectively. Unfortunately, one does not always succeed in deriving
analytic expressions which are of this restrictive form.

In \cite{FKS} we obtained 
\begin{eqnarray}
&&P_{l}(\psi )=1-\Bigg\{ \int\limits_{0}^{\tau }\Vert (V(\vec{x}%
-c(t)e_{z})-V(\vec{x}))\psi \Vert dt  \nonumber \\
&&+\frac{2}{2E+b(\tau )^{2}}\Vert (V(\vec{x}-c(\tau )e_{z})-V(\vec{x}))\psi
\Vert  \nonumber \\
&&+\frac{2|b(\tau )|}{2E+b(\tau )^{2}}\Vert p_{z}\psi \Vert \Bigg\}^{2},
\end{eqnarray}
which is valid when $-E<b(\tau )^{2}/2.$ Here $E$ is the binding energy.
With the same restriction on $b(\tau )$ we found as an upper bound 
\begin{eqnarray}
P_{u}(\psi ) &=&\Bigg\{ \int\limits_{0}^{\tau }\Vert (V(\vec{x}-c(t)e_{z})-V(%
\vec{x}))\psi \Vert dt  \label{Bgen} \\
&&+|c(\tau )|~\Vert p_{z}\psi \Vert +\frac{2|b(\tau )|}{2E+b(\tau )^{2}}%
\Vert p_{z}\psi \Vert \Bigg\}^{2}\;.\;  \nonumber
\end{eqnarray}
By a slightly different analysis we also derived a bound valid without any
additional restrictions 
\begin{eqnarray}
P_{u}(\psi ) &=&\Bigg\{ \int\limits_{0}^{\tau }\Vert (V(\vec{x}-c(t)e_{z})-V(%
\vec{x}))\psi \Vert dt  \nonumber \\
&&+|c(\tau )|~\Vert p_{z}\psi \Vert +|b(\tau )|~\Vert z\psi \Vert \Bigg\}%
^{2}\;.  \label{Hgenu}
\end{eqnarray}
In \cite{FFS} we applied these bounds to the Hydrogen atom, obtaining 
\begin{eqnarray}
&&\!\!\!\!\!\!\!\!\!\!\!\!\!\!\!\!\!\!P_{l}(\psi _{n00})=1-\Bigg\{ \frac{2}{%
n^{3/2}}\tau +\frac{4}{b(\tau )^{2}-1/n^{2}}\frac{1}{n^{3/2}}  \nonumber \\
&&\qquad \qquad \qquad +\frac{1}{n\sqrt{3}}\frac{2|b(\tau )|}{b(\tau
)^{2}-1/n^{2}}\Bigg\}^{2}  \label{HB1} \\
&&\!\!\!\!\!\!\!\!\!\!\!\!\!\!\!\!\!\!P_{u}(\psi _{n00})=\Bigg\{ \frac{2\tau 
}{n^{3/2}}+\frac{|c(\tau )|}{n\sqrt{3}}+\sqrt{\frac{5n^{3}+n}{6}}|b(\tau )|%
\Bigg\}^{2}\,.  \label{HB2}
\end{eqnarray}
Our method allows in principle to consider any bound state, but initially we
restricted ourselves to s-wave functions, keeping however the dependence on
the principal quantum number $n.$ The discussion of (\ref{Bgen}) and (\ref
{Hgenu}) in \cite{FKS} was plagued by the requirement that the pulse
duration should be fairly small. This limitation, which ensured that bounds
for the ionization probability are between zero and one, was overcome in 
\cite{FFS}, since in there we had an additional parameter, i.e. $n$, at
hand. As the expressions (\ref{HB1}) and (\ref{HB2}) show after a quick
inspection, one may achieve that their values remain physical, even if one
increases the pulse duration, but now together with $n$.

In particular we investigated the effect resulting from different pulse
shapes, since it is widely claimed in the literature that a necessary
condition for the existence of stabilization is an adiabatically smooth turn
on (sometimes also off) of the laser field. For definiteness we assumed the
laser light to be of the general form $E(t)=E_{0}\sin (\omega t)g(t).$
Besides other pulses we investigated in particular the ones which are widely
used in the literature, where the enveloping function is either trapezoidal

\begin{equation}
g(t)=\left\{ 
\begin{array}{ll}
\frac{t}{T} & \hbox{for}\qquad 0\leq t\leq T \\ 
1 & \hbox{for}\qquad T<t<(\tau -T) \\ 
\frac{(\tau -t)}{T} & \hbox{for}\qquad (\tau -T)\leq t\leq \tau
\end{array}
\right.  \label{trap}
\end{equation}
or of sine-squared shape 
\begin{equation}
\tilde{g}(t)=\left\{ 
\begin{array}{ll}
\sin ^{2}\left( \frac{\pi t}{2T}\right) & \hbox{for}\qquad 0\leq t\leq T \\ 
1 & \hbox{for}\qquad T<t<(\tau -T) \\ 
\sin ^{2}\left( \frac{\pi (\tau -t)}{2T}\right) & \hbox{for}\qquad (\tau
-T)\leq t\leq \tau .
\end{array}
\right.  \label{sine}
\end{equation}
We found no evidence for stabilization for an \textit{adiabatically smoothly
switched on field}, provided that $b(\tau )\neq 0$. The latter restriction
emerges in our analysis as a technical requirement, but it will turn out
that it is of a deeper physical nature.\footnote{%
Already in (\ref{limh}) we observed that $b(\tau )=0$ is somewhat special.}

\section{Ionization Probability in the ultra-extreme Intensity Limit}

Since we are interested in very high intensities we expect to be able to
draw some conclusions from the expressions for the ionization probability in
which the field amplitude is taken to its ultra-extreme limit, i.e.
infinity. In particular we may decide whether asymptotic stabilization
exists. Despite the fact that in this regime one should commence with a
relativistic treatment, our physical framework, that is the Schr\"{o}dinger
equation (\ref{Schrodi}) remains self-consistent, and should certainly
represent the overall behaviour. In \cite{FKS2} we rigorously take this
limit under certain general assumptions on the atomic potential\footnote{%
We made the general assumptions that $V\left( \vec{x}\right) $\ is a real
measurable function on $R^{3}$ and that \ each $\varepsilon >0$ \ one may
decompose $V$\ as $V=V_{1}+V_{2}$ where $V_{1}$\ is in $L^{2}\left(
R^{3}\right) $ \ (i.e. square integrable) with compact support and $V_{2}$\
\ is in $L^{\infty }\left( R^{3}\right) $\ with $\left\| V_{2}\right\|
_{\infty }=$ess$\sup\limits_{\vec{x}\in R^{3}}\left| V_{2}(\vec{x})\right|
\leq \varepsilon .$ Furthermore we assumed that $H_{l}^{A}$ has no positive
bound states. Such potentials are Kato small. In particular the Coulomb
potential is Kato small.} and the laser field, which include almost all
physical situations discussed in the literature. Whenever $b(\tau )$ and $%
c(\tau )$ vanish simultaneously we found 
\begin{equation}
\lim_{\left| E_{0}\right| \rightarrow \infty }\mathcal{P}\left( \psi \right)
=\left\| e^{-i\tau H_{l}^{0}}\psi \right\| \leq 1\quad  \label{Ex1}
\end{equation}
whereas in all other cases we obtained 
\begin{equation}
\lim_{\left| E_{0}\right| \rightarrow \infty }\mathcal{P}\left( \psi \right)
=1\quad .  \label{Ex2}
\end{equation}
This means that in the former case we have asymptotically weak
stabilization. It should be noted that in this analysis the pulse shape is
kept fixed, such that adiabaticity can not be guaranteed anymore. Hence,
weak stabilization is found for a situation in which it is generally not
expected to occur. It would be very interesting to perform similar
computations as in \cite{FKS2} in which the pulse shape is varied in order
to keep also adiabaticity and study whether the effect will become enhanced
in any way. Furthermore, in this case the time evolution operator coincides
with the one of the free particle $H_{l}^{0}$%
\begin{equation}
\lim_{|E_{0}|\rightarrow \infty }\left\| \left( U_{l}^{S}(\tau ,0)-\exp
-i\tau H_{l}^{0}\right) \psi \right\| =0.
\end{equation}
Of course this type of argument does not allow to draw any decisive
conclusions concerning strong stabilization.

\section{Gordon-Volkov Perturbation Theory}

In the high intensity regime for the radiation fields, the basic assumption
for the validity of conventional perturbation theory breaks down, i.e. that
the absolute value of the potential is large in comparison with the absolute
value of the field. However, there is a replacement for this, the so-called
Gordon-Volkov (GV) perturbation theory \cite{GordonVolkov}. Since the basic
idea is simple, it makes this approach very attractive. Instead of
constructing the power series, either for the fields or for the time
evolution operator, out of the solution for the Schr\"{o}dinger equation
involving the Hamiltonian $H_{l}^{A}$ and regarding\textbf{\ }$\mathbf{z}%
E(t) $ as the perturbation, one constructs the series out of solutions
involving the Hamiltonian $H_{l}^{GV}$ and treats the potential $V$ as the
perturbation.

The starting point in this analysis is the Du Hamel formula, which gives a
relation between two time evolution operators $U_{i}^{a}(t,t^{\prime })$ and 
$U_{j}^{b}(t,t^{\prime })$ associated to two different Hamiltonians $%
H_{i}^{a}(t)$ and $H_{j}^{b}(t)$, respectively 
\begin{equation}
U_{i}^{a}(t,t^{\prime })=U_{j}^{b}(t,t^{\prime
})\!-\!i\!\!\!\int\limits_{t^{\prime }}^{t}ds\,\,U_{i}^{a}\left( t,s\right)
H_{i,j}^{a,b}(s)U_{j}^{b}(s,t^{\prime }).  \label{DUHA}
\end{equation}
Here we use the notation $H_{i,j}^{a,b}(s)=H_{i}^{a}(s)-H_{j}^{b}(s)$. The
formal iteration of (\ref{DUHA}) yields the perturbative series 
\begin{equation}
U_{i}^{a}(t,t^{\prime }) =\sum\limits_{n=0}^{\infty
}U_{i,j}^{a,b}(n|t,t^{\prime }) \quad .  \label{pertpower}
\end{equation}
We introduced in an obvious notation the quantity $U_{i,j}^{a,b}(n|t,t^{%
\prime })$ relating to the time evolution operator order by order in
perturbation theory, i.e. $U_{i,j}^{a,b}(0|t,t^{\prime
})=U_{j}^{b}(t,t^{\prime }),$ $U_{i,j}^{a,b}(1|t,t^{\prime
})=i\int_{t^{\prime }}^{t}\!ds\,\,U_{j}^{b}\left( t,s\right)
H_{i,j}^{a,b}(s)U_{j}^{b}(s,t^{\prime })$, etc. It should be noted that the
perturbative series is gauge invariant in each order, since $%
U_{i,j}^{a,b}(n|t,t^{\prime })$ is a gauge invariant quantity by itself.
Mixing however expansions for different choices of the Hamiltonians, i.e. $a$
and $b$ or different gauges $i$ and $j$ will not guarantee this property in
general. A rather unnatural choice (for instance with regard to the possible
convergence of the series) would be $i\neq j$. Taking therefore $i=j$ and in
addition $a=S$ and $b=GV$ we obtain

\begin{equation}
U_{i}^{S}(t,t^{^{\prime }})=U_{i}^{GV}(t,t^{^{\prime
}})+U_{i,i}^{S,GV}(1|t,t^{\prime })+\ldots \quad  \label{DHGVV}
\end{equation}

\noindent In this case we need $H_{l,l}^{S,GV}(t)=$ $H_{v,v}^{S,GV}(t)=V%
\left( \mathbf{\vec{x}}\right) ,$ $H_{KH,KH}^{S,GV}(t)=V(\mathbf{\vec{x}}%
-c(t)e_{z})$ and the Gordon-Volkov time evolution operator, which in the
KH-gauge equals the free-particle evolution operator in the length gauge 
\begin{eqnarray}
U_{KH}^{GV}(t,t^{\prime }) &=&A_{l\leftarrow
KH}^{-1}(t)U_{l}^{GV}(t,t^{\prime })A_{l\leftarrow KH}(t^{\prime }) 
\nonumber \\
&=&U_{l}^{0}(t,t^{\prime }).  \label{GVKH}
\end{eqnarray}
The expressions for the Gordon-Volkov time evolution operator in the length
and velocity gauge may then simply be obtained from (\ref{GVKH}) by the
application of (\ref{g2})-(\ref{g3}) according to (\ref{Ugauge}). The choice 
$i=j$ together with $a=S$ and $b=A$ in (\ref{DUHA}) yields the usual
perturbation series, which is well known from the low intensity regime. One
may also decide for a rather strange procedure and take the latter choice in
the first iterative step and terminate the series after the second iterative
step in which one makes the former choice. In that case one obtains 
\begin{eqnarray}
&&U_{i}^{S}(t,t^{^{\prime }})=U_{i}^{A}(t,t^{^{\prime }}) \\
&&-i\int\limits_{t^{\prime }}^{t}ds\,\,U_{i}^{A}\left( t,s\right)
H_{i,i}^{S,A}(s)U_{i}^{GV}(s,t^{\prime })+\mathcal{O}(n^{2}).\quad  \nonumber
\end{eqnarray}
For $i=l$ or $i=v$, this procedure is sometimes referred to as the Keldysh- 
\cite{Kel} or Faisal-Reiss \cite{FR} approximation, respectively. As we
demonstrated this method is of course not ``non-perturbative'', as sometimes
wrongly stated in the literature.

There are some exact results which may be derived from the perturbative
expression, one concerning the ultra-extreme intensity limit of the previous
section and the other the ultra-extreme high frequency limit. Both results
are simple consequences of the Riemann-Lebesgue theorem\footnote{%
If $g(x)\in L_{1}(-\infty ,\infty )$ (i.e. $\left| g(x)\right| $ is
integrable) then $\lim\limits_{t\rightarrow \pm \infty }\int\limits_{-\infty
}^{\infty }g(x)e^{-itx}dx=0$.}. We obtain 
\begin{equation}
\lim\limits_{\omega \rightarrow \infty }A_{i\leftarrow j}(t)=1
\end{equation}
such that with (\ref{GVKH}) 
\begin{equation}
\lim\limits_{\omega \rightarrow \infty }U_{KH}^{GV}(t,t^{\prime
})=U_{l}^{0}(t,t^{\prime })=e^{-i(t-t^{\prime })H_{l}^{0}}.  \label{Lim}
\end{equation}
Since the atomic potential is independent of $E_{0}$ we obtain with (\ref
{Lim}) that the entire Gordon-Volkov series (\ref{pertpower}) is independent
of the field amplitude as well, such that 
\begin{equation}
\frac{d}{dE_{0}}\left( \lim\limits_{\omega \rightarrow \infty }\mathcal{P}%
(\psi )\right) =0 \quad .
\end{equation}

We have therefore weak stabilization in this ultra-extreme high frequency
limit for all systems for which (\ref{pertpower}) makes sense and for which
the laser field is of the form $E(t)=E_{0}\sin (\omega t)g(t)$. One should
keep in mind that the use of the Stark Hamiltonian in (\ref{Schrodi})
assumes the validity of the dipole approximation, such that the limit $%
\omega \rightarrow \infty $ (a slightly milder assumption was used in the
seminal paper \cite{Gavrila} which formulates the high frequency approach)
only makes formally sense. In order to describe real physics in this
frequency regime one should actually also take multipole terms into account.
This is a further example of the errors we mentioned at the beginning of
section 3.

Concerning the ultra-extreme intensity limit, we consider the transition
amplitude between two bound states $\psi _{i}(\vec{x}),$ $\psi _{j}(\vec{x})$
of the Hamiltonian $H_{l}^{A}$ perturbatively 
\begin{eqnarray}
&& \!\!\!\!\! \left\langle \psi _{i},U_{l}^{S}(\tau ,0)\psi _{j}
\right\rangle  \label{Expert} \\
&& \!\!\!\!\! =\left\langle \psi _{i},A_{KH\leftarrow l}(\tau
)U_{KH}^{S}(\tau ,0)\psi _{j}\right\rangle  \nonumber \\
&& \!\!\!\!\! =\left\langle \psi _{i},A_{KH\leftarrow l} (\tau
)U_{KH}^{GV}(\tau ,0)\psi _{j}\right\rangle  \nonumber \\
&&\!\!\!\!\! +\left\langle \psi _{i},A_{KH\leftarrow l}(\tau
)U_{KH,KH}^{S,GV}(1|\tau ,0)\psi _{j}\right\rangle +\ldots  \nonumber
\end{eqnarray}
Recall that $a(0)=b(0)=c(0)=0,$ such that $A_{l\leftarrow KH}(0)=1$. Using
now (\ref{GVKH}) it is clear that to zeroth order we obtain 
\begin{equation}
\left\langle \psi _{i}(\vec{x}),e^{-i\tau H_{l}^{0}}\psi _{j}(\vec{x}%
)\right\rangle \quad ,
\end{equation}
when $b(\tau )=c(\tau )=0.$ In all other cases we may bring this term into a
form suitable for the application of the Riemann-Lebesgue theorem, such that
the zeroth order matrix element always vanishes in the ultra-extreme
intensity limit. For the higher order terms the argument is analogous with
the difference that the condition $b(\tau )=c(\tau )=0$ does not have the
consequence that these expressions become independent of $E_{0},$ since also
terms like $b(t),c(t)$ for $0<t<\tau $ appear. Hence by the application of
the Riemann-Lebesgue theorem all higher order terms vanish in the limit $%
E_{0}\rightarrow \infty .$ If we now sum over all bound states $i$ in (\ref
{Expert}) we obtain the results of section 4 (\ref{Ex1}) and (\ref{Ex2}).

\section{1-dimensional $\delta -$potential}

In \cite{FFS2} we applied the GV-perturbation theory to the one-dimensional
delta potential with coupling constant $\alpha $%
\begin{equation}
V\left( \mathbf{x}\right) =-\alpha \,\delta \left( \mathbf{x}\right)
\,.\quad \,\smallskip
\end{equation}
In the momentum space representation this potential becomes 
\begin{equation}
V(p,p^{\prime })=\left\langle p\left| V\right| p^{\prime }\right\rangle =-%
\frac{\alpha }{2\pi }
\end{equation}
and the wave function for its only bound-state is well known to be 
\begin{equation}
\psi (p,t=0)=\sqrt{\frac{2}{\pi }}\frac{\alpha ^{\frac{3}{2}}}{\alpha
^{2}+p^{2}}.
\end{equation}
The very fact that this potential possesses one bound state only with bound
state energy $-\alpha ^{2}/2$ makes it a very attractive theoretical atomic
toy potential (e.g. \cite{delta,Eb1,Olg2,Beck}). Also with regard to the
GV-perturbation theory one expects intuitively a good convergence.

We construct the exact time-dependent wave function: 
\begin{equation}
\psi \left( p,t\right) =\psi _{GV}\left( p,t\right) +\Psi \left( p,t\right)
\label{bas}
\end{equation}
with 
\begin{eqnarray}
\Psi \left( p,t\right) &=&i\frac{\alpha }{2\pi }\int%
\limits_{0}^{t}dse^{-ia(t)}e^{ic_{ts}(p-b(t))}  \nonumber \\
&&\times e^{-\frac{i}{2}(p-b(t))^{2}(t-s)}\psi _{I}\left( s\right) \\
\psi _{I}\left( t\right) &=&e^{ia(t)}\int\limits_{-\infty }^{\infty
}dp\,\psi \left( p,t\right) \quad \,.
\end{eqnarray}
Integrating (\ref{bas}) with respect to $p$ we obtain a Volterra equation of
the second kind in $t$%
\begin{equation}
\psi _{I}\left( t\right) =\int\limits_{-\infty }^{\infty }dp\,\psi
_{GV}\left( p,t\right) +\sqrt{\frac{i\alpha ^{2}}{2\pi }}\int%
\limits_{0}^{t}ds\psi _{I}\left( s\right) \frac{e^{i\frac{c_{ts}^{2}}{2(t-s)}%
}}{\sqrt{t-s}}.  \label{Volt}
\end{equation}
The virtue of this equation is that the error of its solution, even when
obtained by an iterative procedure, is completely controllable. The
iteration of the Volterra equation yields 
\begin{equation}
\psi \left( t\right) =\int\limits_{-\infty }^{\infty }dp\,\psi _{GV}\left(
p,t\right) +\sum\limits_{n=1}^{\infty }\psi _{n}\left( t\right)
\end{equation}
with $\psi _{n}\left( t\right) $ denoting the function order by order. We
derived \cite{FFS2} an upper bound for the absolute value of this function 
\begin{equation}
\left| \psi _{n}\left( t\right) \right| \leq \sqrt{8\alpha \pi }\frac{1}{%
n\Gamma \left( \frac{n}{2}\right) }\left( \alpha \sqrt{t/2}\right) ^{n}.
\end{equation}
It is known from the theory of integral equations \cite{Intequ} that this is
sufficient to prove that the series converges for \underline{all} values of $%
\alpha$, and in addition we were even able to sum up the whole series 
\begin{eqnarray}
\sum\limits_{n=1}^{\infty }\left| \psi _{n}\left( t\right) \right| &=&\sqrt{%
2\alpha \pi }\left( 2\exp (\alpha ^{2}t/2)-1\right. \,  \nonumber \\
&&\left. -U_{\frac{1}{2},\frac{1}{2}}\left( \alpha ^{2}t/2\right) /\sqrt{\pi 
}\right)
\end{eqnarray}
such that we can compute the maximal relative error after the zeroth order
in GV perturbation theory 
\begin{equation}
\mu =2\sqrt{\pi }\left| \frac{\left( 2\exp (\alpha ^{2}t/2)-1-U_{\frac{1}{2},%
\frac{1}{2}}\left( \alpha ^{2}t/2\right) /\sqrt{\pi }\right) }{\left( U_{%
\frac{1}{2},\frac{1}{2}}\left( \Phi _{-}\right) +U_{\frac{1}{2},\frac{1}{2}%
}\left( \Phi _{+}\right) \right) }\right| .
\end{equation}
\noindent Here $U_{\frac{1}{2},\frac{1}{2}}\left( z\right) $ is the
confluent hypergeometric function (see for instance \cite{Abr}) and 
\begin{equation}
\Phi _{\pm }:=\tau \alpha ^{2}\left( \pm \gamma +\frac{i}{2}\left( 1-\gamma
^{2}\right) \right) ,\gamma :=\frac{c(\tau )}{\tau \alpha }.
\end{equation}
This analysis allows us to determine the error which is introduced by the
termination of the GV-series. Usually this is done without any justification
about the precision and the only reasoning provided is very often solely the
comparison with the next order term. This is however not enough as one knows
from simple rest term estimations in a Taylor expansion.

The ionization probability turns out to be

\begin{eqnarray}
\mathcal{P}\left( \psi \right) &=&1-\left| q(\tau )\right| ^{2} \\
&=&1-\left| \left\langle \psi ,\psi _{GV}(\tau )\right\rangle +\left\langle
\psi ,\Psi (\tau )\right\rangle \right| ^{2}\quad
\end{eqnarray}
with 
\begin{eqnarray}
&&\left\langle \psi ,\psi _{GV}(\tau )\right\rangle =\frac{2}{\pi }\alpha
^{3}e^{-ia(\tau )} \\
&&\qquad \quad \times \int\limits_{-\infty }^{\infty }dp\frac{\exp \left(
-i\tau \frac{p^{2}}{2}-ic\left( \tau \right) p\right) }{\left( \alpha
^{2}+\left( p+b\left( \tau \right) \right) ^{2}\right) \left( \alpha
^{2}+p^{2}\right) }  \nonumber
\end{eqnarray}
and 
\begin{eqnarray}
&&\left\langle \psi ,\Psi (\tau )\right\rangle =ie^{-ia(\tau )}\sqrt{\frac{%
\alpha ^{5}}{2\pi ^{3}}}\int\limits_{0}^{\tau }ds\psi _{I}\left( s\right) \\
&&\qquad \quad \times \int\limits_{-\infty }^{\infty }dp\frac{e^{ic_{\tau
s}(p-b(\tau ))}e^{-\frac{i}{2}(p-b(\tau ))^{2}(\tau -s)}}{\left( \alpha
^{2}+p^{2}\right) }\,\quad .  \nonumber
\end{eqnarray}
$\left| q(\tau )\right| ^{2}$ has of course the interpretation as survival
probability. We use the abreviation $c_{tt^{\prime }}:=c(t)-c(t^{\prime }).$
These expressions constitute explicit examples for the general statements
made at the end of section 5. Taking the ultra-extreme intensity limit we
obtain as a consequence of the Riemann-Lebesgue theorem 
\begin{equation}
\lim_{E_{0}\rightarrow \infty }\left| q(\tau )\right| ^{2}=\QATOPD\{ .
{\left| \left\langle \psi ,\psi _{GV}(\tau )\right\rangle \right| ^{2}\,%
\text{for }b(\tau )=c(\tau )=0}{0\qquad \qquad \qquad \text{otherwise}}
\label{ultra}
\end{equation}
This means we have asymptotically weak stabilization and the result is in
agreement with the one in section 4. However, the assumption made on the
potential in \cite{FKS2} does not include the $\delta $-potential\footnote{%
The $\delta $-potential is not a Kato small potential.}, such that (\ref
{ultra}) is not only obtained by an alternative method but also covers an
additional case. As we already observed, a different physical behaviour is
obtained depending on the values of $b(\tau )$ and $c(\tau )$ and it is
therefore instructive to treat several cases separately: To the lowest order
we obtained

\begin{itemize}
\item[i)]  $b(\tau )=0,$ $c(\tau )=0\medskip $%
\begin{equation}
\mathcal{P}\left( \psi \right) =1-\frac{4}{\pi }\left| U_{\frac{1}{2},-\frac{%
1}{2}}\left( i\tau \frac{\alpha ^{2}}{2}\right) \right| ^{2}\,\quad
\end{equation}

\item[ii)]  $b(\tau )=0,$\emph{\ }$c(\tau )\neq 0\medskip $%
\begin{eqnarray}
q(\tau )=\varphi _{-}\,U_{\frac{1}{2},\frac{1}{2}}\left( \Phi _{+}\right)
+\varphi _{+\,}U_{\frac{1}{2},\frac{1}{2}}\left( \Phi _{-}\right)  \nonumber
\\
+i\tau \alpha ^{2}U_{\frac{1}{2},\frac{3}{2}}\left( \frac{i\tau \alpha ^{2}}{%
2}\right) \,\,  \label{ii}
\end{eqnarray}
\smallskip with $\varphi _{\pm }=\frac{1}{2}\left( 1\pm \alpha c(\tau
)-i\tau \alpha ^{2}\right) $\smallskip . Equation (\ref{ii}) is only valid
for $\left| \gamma \right| <1$.

\item[iii)]  $b(\tau )\neq 0,$ $c(\tau )\neq 0$%
\begin{eqnarray}
q(\tau )=\frac{\tilde{\varphi}_{+}}{\sqrt{\pi }}\left[ U_{\frac{1}{2},\frac{1%
}{2}}\left( \Phi _{+}\right) +U_{\frac{1}{2},\frac{1}{2}}(\tilde{\Phi}%
_{-})\right]  \label{iii} \\
+\frac{\tilde{\varphi}_{-}}{\sqrt{\pi }}\left[ U_{\frac{1}{2},\frac{1}{2}%
}\left( \Phi _{-}\right) +U_{\frac{1}{2},\frac{1}{2}}(\tilde{\Phi}%
_{+})\right]  \nonumber
\end{eqnarray}
\end{itemize}

with 
\begin{eqnarray}
\tilde{\varphi}_{\pm } &=&\frac{\alpha ^{2}}{b(\tau )^{2}\pm 2\alpha b(\tau
)i}\quad \\
\quad \tilde{\Phi}_{\pm } &=&\tau \alpha ^{2}\left( \pm \tilde{\gamma}+\frac{%
i}{2}\left( 1-\tilde{\gamma}^{2}\right) \right)
\end{eqnarray}

Equation (\ref{iii}) is only valid for $\left| \tilde{\gamma}\right| =\left| 
\frac{c(\tau )}{\tau \alpha }+\frac{b(\tau )}{\alpha }\right| <1.$

The restrictions on the parameters $\gamma $ and $\tilde{\gamma}$ originate
in the limited validity of the integral representation for the confluent
hypergeometric functions, which is employed here. To obtain all higher
orders we have to compute 
\begin{eqnarray}
&&\left\langle \psi ,\Psi \left( \tau \right) \right\rangle =\frac{i\alpha ^{%
\frac{3}{2}}e^{-ia(\tau )}}{2\pi \sqrt{2}}  \label{Higher} \\
&&\int\limits_{0}^{\tau }ds\psi _{I}\left( s\right) e^{\frac{ic_{\tau s}}{%
2(\tau -s)}}\left( U_{\frac{1}{2},\frac{1}{2}}\left( \Phi _{-}^{\prime
}\right) +U_{\frac{1}{2},\frac{1}{2}}\left( \Phi _{+}^{\prime }\right)
\right)  \nonumber
\end{eqnarray}
with 
\begin{eqnarray}
\!\!\!\!\!\!\!\!\!\!\!\!\Phi _{\pm }^{\prime } &=&\pm \alpha \left(
c_{ts}-\left( t-s\right) b\left( t\right) \right)  \nonumber \\
&&+\frac{i}{2}\left( t-s\right) \left( \alpha ^{2}-\left( \frac{c_{ts}}{t-s}%
-b\left( t\right) \right) ^{2}\right) .\,
\end{eqnarray}
Hence the problem to compute the ionization probability has been reduced to
solving (\ref{Volt}) and subsequently evaluate (\ref{Higher}). Surely this
is not possible to perform entirely in an analytical way, but the initial
problem has now been reduced to a numerical task, whose error is well under
control. In \cite{FFS2} we carried out this analysis for a pulse involving
the trapezoidal enveloping function. We do not find any evidence for strong
stabilization even for $b(\tau )=0$ and $c(\tau )=0,$ however,
asymptotically weak stabilization exists for the latter case.

\section{Conclusions}

The main outcome of our investigations is that the governing parameters for
the behaviour of an atom in an intense laser field are the total classical
momentum transfer $b(\tau )$ and the total classical displacement $c(\tau )%
\footnote{%
Some authors claim \cite{Eberly} that for a pulse of the form (\ref{trap}) $%
b(T)$ and $c(T)$ should be relevant parameters. Besides the fact that for a
smoothly differentiable pulse (e.g. (\ref{sine})) these quantities are not
precisely defined they do not emerge in our analysis as significant.}$.
Whenever both these two quantities vanish, asymptotically weak stabilization
does exist. Also the authors of \cite{Pot} found asymptotically weak
stabilization for this situation (see Fig. 3 therein). For all other cases
we did not find any evidence for stabilization.

Since our findings apparently differ from many theoretical results other
authors obtained by alternative methods we would like comment on possible
resolutions for this discrepancy:

Introducing a cut-off in the number of bound states will produce an upper
bound for the ionization probability, since in this way one effectively
enlarges the continuum. In case one finds stabilization for such a bound one
could confidently conclude that this effect indeed exists. Since for lower
intensities one can certainly expect this bound to be relatively close to
the real value, whereas for higher intensities this bound should decrease
even more.

The introduction of a cut-off (e.g. equation (3) in \cite{SuEb}) in the
``discrete continuum'' (besides the fact that this is an ill-defined
concept) yields a lower bound for the real ionization probability, which is
expected to be relatively accurate for low intensities but very far from the
real value for high intensities. Hence keeping the cut-off fixed and
interpreting the result obtained in this way as ``exact'', one has certainly
introduced an artificial mechanism to ``create'' stabilization.

A further approach is based on the Fourier expansion of the Hamiltonian in
the KH-gauge and thereafter simply keeping the lowest terms. The way this,
in principle legitimate, method is carried out sometimes makes conceptual
assumptions which are in clear conflict to our main physical framework. For
instance the basic Hamiltonian used in \cite{Olg} (equation (1) therein) is
in our notation $H_{KH}^{S}(t)$ for an instantaneously switched on
monochromatic laser pulse, i.e. $E(t)=E_{0}\cos (\omega t)$. However, the
authors of \cite{Olg} claim stabilization to exist for pulses with a smooth
adiabatic turn on and off (gau\ss ian enveloping function). Clearly equation
(6) in \cite{Olg} breaks the gauge invariance discussed in section 2.1. of
our manuscript, such that the authors consider an entirely different system.
In other words the Hamiltonian (6) in \cite{Olg} is not gauge equivalent to $%
H_{l}^{S}(t)$ for a laser pulse with gau\ss ian enveloping function. The
potential in $H_{KH}^{S}(t)$ is shifted by $c(t)$ and not $E(t)$. From our
point of view the findings of the authors are not surprising since they
artificially impose that $c(\tau )=0$, (we also find asymptotically weak
stabilization in this case) such that one should solve the inverse problem
in this case to find out which pulse is really considered there. However,
even under these assumptions we would still not find strong stabilization.
The same procedure is used for instance in \cite{Olg2,Olg3}.

Concerning the investigations which do not find any evidence for
stabilization at all, we would like to make the following comment on the use
of the GV-perturbation theory. In the last reference of \cite{Gelt} a pulse
with instantaneous switch on was used in this context, i.e. $E(t)=E_{0}\cos
(\omega t),$ and an analysis up to first order GV-perturbation theory was
carried out. Typical parameters in \cite{Gelt} were $\alpha
=1/2,E_{0}=5,\omega =1.5$ and the pulse length was 2 cycles, that is $\tau
\sim 8$. For these parameters we obtain for the relative maximal error $\mu
\approx 8.44$, such that we do not expect the GV-perturbation series to be a
good approximation up to this order and statements made in this context
should be treated with extreme care.

We would like to conclude with a remark on the existing experimental
findings \cite{Mul}. So far the experiments carried out only find evidence
for asymptotically weak stabilization and confirmations for the existence of
strong stabilization do not exist to our knowledge. We are therefore in
complete qualitative (it is difficult to determine which values $b(\tau )$
and $c(\tau )$ have for the experimentally employed pulses (see also
footnote 3 for this)) agreement with the existing experiments.

\textbf{Acknowledgment:} We would like to thank H.G. Muller and R. Sauerbrey
for useful discussions and comments. CFMF is supported by the DAAD. A.F. and
R.S. are grateful to the Deutsche Forschungsgemeinschaft (Sfb288) for
partial support.

\end{document}